\documentclass{ws-procs9x6}

\begin{document}

\title{  One-loop contributions of charginos and neutralinos to 
$W^+W^-$ production in $e^+e^-$ 
collisions}

\author{Kaoru Hagiwara}

\address{Theory group, KEK, Tsukuba, Ibaraki, 305-0851,Japan \\
E-mail: kaoru.hagiwara@kek.jp}

\author{Shinya Kanemura}

\address{Department of Physics, Osaka University, 
Toyonaka, Osaka 560-0043, Japan\\
E-mail: kanemu@het.phys.sci.osaka-u.ac.jp}

\author{Yoshiaki Umeda\footnote{\uppercase{S}upported by 
\uppercase{BMBF}, contract nr. 05 
\uppercase{HT1PAA} 4 and 
\uppercase{NSC} 92-2811-\uppercase{M}-009-029.}}

\address{Institute of physics, National Chiao Tung University,
Hsinchu 300, Taiwan\\
E-mail: umeda@faculty.nctu.edu.tw}


\maketitle

\abstracts{
We study one-loop effects of charginos and neutralinos on the 
helicity amplitudes for the process $e^+e^-\rightarrow W^+W^-$ 
in the MSSM.
The calculation is tested by using two methods.  
First, the sum rules among form factors 
which result from BRS invariance.
Second, the decoupling property in the large mass limit is used 
to test the overall normalization of the amplitudes. 
The one-loop contribution to the helicity amplitudes 
is shown as a function of
$\mu$ Higgs parameter and $\sqrt{s}$.
We also study the effects of the $CP$-violating phase. 
}
\section{Introduction}
We discuss the one-loop charginos and neutralinos
contributions to the $e^+(k,\tau)e^-(\overline{k},\overline{\tau})
\rightarrow W^+(p,\lambda) W^- (\overline{k},\overline{\lambda})$
in the MSSM\cite{ino}. Here $k$ and $p$ are the momentum, 
and $\tau$ and $\lambda$
denote helicities. 
The mass matrix of charginos or neutralinos
are express by 2 $\times$ 2 or $ 4\times 4$ matrix, respectively.
The helicity amplitudes can be written by using 16 basis tensors
$T_i^{\mu\alpha\beta}$
\begin{equation}
\label{amp-eeww}
{M}^{\lambda \overline{\lambda}}_\tau
= \sum_{i=1}^{16} F_{i,\tau}(s,t)\, j_\mu(k,\overline{k},\tau) 
T_i^{\mu\alpha\beta} \epsilon_\alpha(p,\lambda)^\ast 
\epsilon_\beta(\overline{p},\overline{\lambda})^\ast  \;,
\vspace*{-3pt}
\end{equation}
where $j_{\mu}$ is the electron current and 
$\epsilon_{\alpha}$ is the polarization vectors for $W$.
All the dynamical information is contained in the form factors
$F_{i,\tau}(s,t)$. Among the 16 form factors, physically polarized 
$W$ boson are described by nine form factors.
To test the physical form factors of $F_{i,\tau}(s,t)$(i=1$-$9), 
we calculate unphysical form factors of 
$F_{i,\tau}(s,t)$(i=10$-$13) together
with the form factors of $H_{i,\tau}(s,t)$(i=1$-$4) of 
the helicity amplitudes of $e^+(k,\tau)e^-(\overline{k},\overline{\tau})
\rightarrow \omega^+(p) W^- (\overline{k},\overline{\lambda})$
($\omega^+$: Nambu-Goldstone boson), by using basis tensors 
$S^{\mu\alpha}_i$
\begin{equation}
 {M}^\lambda_\tau (e^+e^-\rightarrow \omega^+ W^-)
 =    i \sum_{i=1}^{4} H_{i,\tau}(s,t)\, j_\mu(k,\overline{k},\tau) 
S_i^{\mu\alpha} \epsilon_\alpha(p,\lambda)^\ast \;.
\label{amp-eewx}
\vspace*{-10pt}
\end{equation}
\section{Test}
In order to ensure the correctness of our calculation,
we examine the BRS invariance of one-loop helicity amplitudes
and the decoupling behavior of the SUSY effects in the large
mass limit of charginos and neutralinos. 

\begin{figure}[t]
\caption{One-loop contributions to $M^{00}_{\tau}$ helicity
amplitudes are shown.}
\centerline{\epsfxsize=2.9in\epsfbox{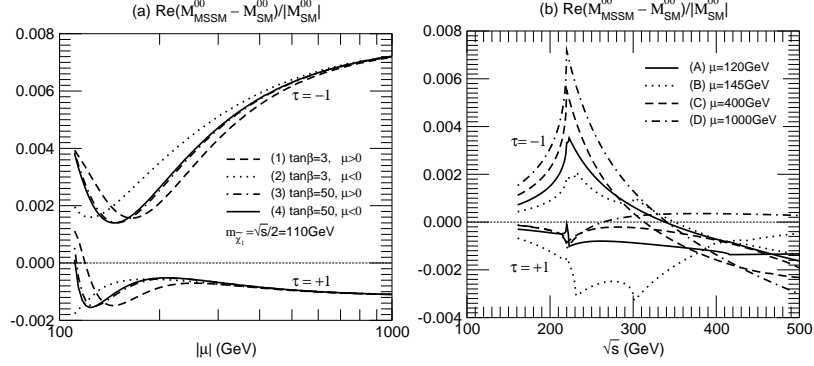}}
\label{fig:m00}
\vspace*{-13pt}
\end{figure}

\subsection{The BRS invariance}
The linear relation between
$e^+e^-\rightarrow W^+_SW^-_P$ and 
$e^+e^-\rightarrow \omega^+W^-_P$ helicity amplitudes
is obtained from BRS invariance,
where index P and S denote physical and scalar 
polarization, respectively.
We obtain three sum rules
among the form factors of $\widetilde{F}_{i,\tau}$ and 
$\widetilde{H}_{3,\tau}$ from BRS invariance\cite{sf}
\begin{equation}
\label{eq:brs2}
\sum^{16}_{j=1} \xi_{i,j}F_{j,\tau}(s,t) = C^{BRS}_{mod}H_{i,\tau}(s,t)\,,
\end{equation}
where i runs from 1 to 3 and $\xi_{i,j}$ are determined by
kinematical parameters. 
For the factor $C_{\rm mod}^{BRS}$, see Ref.~[2].
We numerically confirmed that the Eq.~(\ref{eq:brs2})
agrees more than 11 digit numerically.  
\subsection{Decoupling limit}
The one-loop effects of the SUSY particles should decouple from the low
energy observable in the large mass limit.  
By using this fact, we can test
the overall normalization of the amplitudes.
We check that the differential cross section of the MSSM become 
that of the SM in the large mass limit of the 
gaugino mass $M_1$ and $M_2$, and  $\mu$ higgs term.
See Fig.~1 of Ref.~[1].

\begin{figure}[t]
\caption{The effects of $CP$-violating phases are shown.}
\centerline{\epsfxsize=2.9in\epsfbox{7f.epsi}}
\label{fig:cp}
\vspace*{-13pt}
\end{figure}

\begin{table}[t]
\tbl{Input parameters and mass spectra. The 
$\tan\beta=$3 for all cases.}
{\footnotesize 
\begin{tabular}{l|rrrrr}
   & A & B & C & D & E\\ \hline
\multicolumn{5}{l}{Parameter}           \\ \hline
$\mu$ (GeV)     & +120 & +145 & +400 & +1000 & +130 \\
$M_2$ (GeV)     & 541 & 242 & 125 & 115  & 158 \\
$\varphi_{1}$  &0&   0 &   0 &    0 & $\frac{2}{3}\pi$ \\
$\varphi_{\mu}$&0&   0 &   0 &    0 & $\frac{2}{3}\pi$ \\
\hline  
\multicolumn{5}{l}{Mass spectra (GeV)} \\ \hline
$m_{\widetilde{\chi}^-_1}$  & 110 & 110 & 110 & 110 & 110\\
$m_{\widetilde{\chi}^-_2}$  & 555 & 283 & 420 &1007 & 207\\
$m_{\widetilde{\chi}^0_1}$  &  99 &  81 &  60 &  57 &  75\\
$m_{\widetilde{\chi}^0_2}$  & 123 & 150 & 111 & 110 & 105\\
$m_{\widetilde{\chi}^0_3}$  & 285 & 150 & 403 &1002 & 154\\
$m_{\widetilde{\chi}^0_4}$  & 555 & 285 & 422 &1007 & 205\\
\hline\hline
\end{tabular}\label{tab:set1}}
\vspace*{-13pt}
\end{table}

\section{Numerical results}
We present the one-loop effects to the helicity amplitudes.
We assume the GUT relation $M_1 = 5M_2\hat{s}^2/3\hat{c}^2$,
light chargino mass $m_{\widetilde{\chi}_1^-}$ = 110GeV,
and scattering angle $\theta = 90^\circ$.
\subsection{The chargino and neutralino contribution to $M^{00}_{\tau}$}
Among the helicity amplitudes, the loop correction is the largest
in 00 helicity amplitude.
In Fig.~\ref{fig:m00}(a), we show the $\mu$ dependence for
$\tan\beta$ = 3 or 50.
The loop effects are about 0.7\% at $|\mu |$ = 1000GeV for $\tau =-1$,
whereas they remain small for $\tau =+1$ at around $0.1\%$ level.
In Fig.~\ref{fig:m00}(b), the $\sqrt{s}$ dependence 
for four sets of parameter (sets A to D of Table~\ref{tab:set1})
is shown. 
Sharp peak can be seen for each curve, which correspond to the 
thresholds of pair production of the lightest chargino 
at  $\sqrt{s}= 220$GeV for $\tau =-1$.
The corrections are negative for $\tau =+1$.
\subsection{$CP$-violating effects}
There are new $CP$-violating phases in the MSSM.
In the chargino and neutralino sectors, it arises from
the $\mu$ Higgs mass term $\mu e^{i \varphi_{\mu}}$ and the gaugino mass 
parameters $M_1 e^{i\varphi_1}$ for the most general case
when we take the phase of $M_2$ be 0 by
rephasing. Large $CP$-violating phases in chargino and neutralino
sectors are possible without contradicting the EDM constraints
of electrons and neutrons, if the parameters for sleptons 
and squarks are adjusted.
In Figs.~\ref{fig:cp}, the real part (solid curve) 
and the imaginary part (dotted curve) of $f_4^Z$, $f_6^\gamma$,
and $f_6^Z$ of Ref.~[1,2] are shown as a function of $\sqrt{s}$
for the parameters of set E in Table~\ref{tab:set1}.
These form factors take their maximum or minimum at around
$\varphi_{\mu}=\varphi_{1}=2/3\pi$ or $4/3\pi$.
We can measure the $CP$-violating effects of $f_4$ and $f_6$ 
by the difference of helicity amplitudes $M^{\pm 0}$ and $M^{\mp 0}$.
The $CP$-violating effects in the helicity amplitudes can be
of the order (0.1\%). 
\section{Conclusion}
The correction due to chargino and neutralino contributions to 
$e^+e^-\rightarrow W^+W^-$ helicity amplitudes can be 
of the order (1\%) and the $CP$-violating effects can be
of the order (0.1\%).

\end{document}